\documentclass[prl,twocolumn,amsmath,amssymb]{revtex4-1} 

\usepackage{epsfig,amsmath}
\usepackage{subfigure}
\usepackage{graphicx}
\usepackage{dcolumn}
\usepackage{stmaryrd}
\usepackage{mathrsfs}
\usepackage{pifont}
\usepackage{amsthm}
\usepackage{amssymb}
\usepackage{mathtools}
\usepackage{bm}
\usepackage{latexsym}
\usepackage[colorlinks=true,linkcolor=blue,citecolor=blue]{hyperref}
\usepackage{color}

\theoremstyle{plain}

\def\pra#1{{ Phys.\ Rev. A\/} {\bf#1}}
\def\prb#1{{ Phys.\ Rev. B\/} {\bf#1}}

\def\prl#1{{ Phys.\ Rev.\ Lett.} {\bf#1}}
\def\pr#1{{ Phys.\ Rev.} {\bf#1}}
\def\sci#1{{ Science} {\bf#1}}

\def\rmp#1{{ Rev. \ Mod. \ Phys.} {\bf#1}}

\def\epl#1{{Europhys.\ Lett. }{\bf #1}}

\begin{document}

\title{Symmetry and Exact Solutions of General Spin-Boson Models}

\author{Yifan Sun$^{1,4}$, and Lian-Ao Wu$^{1,2,3}$}\thanks{Author to whom any correspondence should be addressed. Email address: lianao.wu@ehu.es}

\affiliation{$^{1}$Department of Physics, The Basque Country University (EHU/UPV), PO Box 644, 48080 Bilbao, Spain \\ $^{2}$Ikerbasque, Basque Foundation for Science, 48011 Bilbao, Spain \\ $^{3}$ EHU Quantum Center, University of the Basque Country UPV/EHU, Leioa, 48940 Biscay, Spain\\
$^{4}$Beijing Key Laboratory of Quantum Matter State Control and Ultra-Precision Measurement Technology, 100081, Beijing, China}


\begin{abstract}
Spin–boson models are the canonical benchmark for quantum dissipation. We show the symmetry structure of general spin–boson Hamiltonians and obtain their spectra explicitly by exploiting the symmetry. As an illustration of the general case, we numerically demonstrate the exact solution for the two-mode case.
\end{abstract}


\maketitle
{\it Introduction.---}Spin–boson models provide the paradigmatic description of quantum dissipation, capturing a two-level system coupled to a bosonic environment. It reveals how environmental fluctuations suppress coherence and, for sufficiently strong coupling and appropriate spectral structure, can induce a quantum phase transition \cite{Leggett1987}. It reduces to the historically seminal quantum Rabi model when the bosonic bath is restricted to a single harmonic oscillator mode---a model first introduced more than eighty years ago \cite{Rabi1936}. Despite its apparent simplicity, the model exhibits remarkably rich physics \cite{Malekakhlagh2019,Lu2023} and underpins a wide range of quantum phenomena across cavity QED \cite{Brune1996,Raimond2001,Mabuchi2002,Walther2006}, circuit QED \cite{Wallraff2004,Blais2004,Chiorescu2004,Schuster2007,Clarke2008,Hofheinz2009}, trapped ions \cite{Leibfried2003,Pedernales2015}, quantum dots \cite{Hennessy2007}, nano-electromechanical systems \cite{Irish2003,Cleland2004,LaHaye2009}, molecular systems \cite{Albert2012}, and condensed-matter platforms \cite{Irish2007}. It can also be efficiently simulated \cite{Wu2002} in engineered settings such as photonic architectures \cite{Crespi2012} and nuclear magnetic resonance platforms \cite{Wu2024}, providing insight into qubit decoherence \cite{Jing2018} and the simulation of diverse quantum statistical properties \cite{Wu2002-1}. Recent advances in engineered quantum systems have enabled access to coupling regimes ranging from the well-understood Jaynes–Cummings limit \cite{Wallraff2004,Blais2004,Chiorescu2004,Schuster2007,Clarke2008,Hofheinz2009,Jaynes1963}, where the rotating-wave approximation (RWA) applies, to the ultra-strong \cite{Niemczyk2010,
FornDiaz2010} and deep-strong \cite{Yoshihara2017} coupling regimes, which require generalized versions of the RWA \cite{Irish2007,Albert2011} and exhibit qualitatively new dynamical effects.
 
A central theme underlying these developments is the symmetry structure and exact solvability of the quantum Rabi model, the first {\it milestone} exact solution given by D. Braak \cite{Braak2011}. Despite extraordinary progress in computational methods, numerical approaches cannot substitute for exact analytic solutions \cite{Schweber1967,Swain1973,Durst1986,Chen2012,Braak2013,Peng2013,Zhang2014,Duan2015,He2015,Wu2017,Xie2017,Cui2017,Peng2017}. An exact solution is a gemstone and fundamentally tied to the symmetry of the model \cite{Wu2003,Wu2017-2}, providing the most concise, rigorous, and conceptually transparent characterization of its physics. Such insight cannot be reproduced by numerical or data-driven approaches. Moreover, without analytic insight, it is often impossible to determine whether numerical results are fundamentally correct, convergent, or unique, particularly in parameter regimes involving spectral complexity, strong nonlinearities, or quantum phase transitions \cite{Wu2021}. These challenges are especially pronounced in spin–boson models, which lack a classical limit and may exhibit nontrivial critical behavior. For these reasons, exact solutions and symmetry-based analyzes are essential, supplying the mathematical structure needed to ensure the reliability and interpretability of numerical studies.


While the single-mode quantum Rabi model is now well understood, far less is known about its multi-mode generalizations, which form the broader family of spin–boson models. The specific form of the model can be expressed by
\begin{equation}\label{eq:HMSpinBos}    H_M=\sum_{j=1}^{N}\omega_{j}a_{j}^{\dagger}a_{j}+\sigma_x\sum_{k=1}^{N}g_k(a_{k} ^{\dagger}+a_{k})+\Delta\sigma_z,
\end{equation}where $\omega_{j}$ is the energy spectrum of the bosons, $g_{k}$ is the coupling of spin and a boson, and $\Delta$ is half of the energy gap of the spin. $a_{j}^{\dagger}$ ($a_{j}$) is the creator (annihilator) of a boson, and $\sigma_x$ ($\sigma_z$) is the Pauli-$X$ operator (Pauli-$Z$ operator). The model arises naturally when a two-level system interacts with a structured bosonic environment and serve as fundamental building blocks for quantum dissipation \cite{Eberly1980}, open-system dynamics \cite{Vaaranta2025}, and many-body light–matter physics \cite{Chen2022}. However, the presence of multiple bosonic degrees of freedom greatly increases the complexity of the problem, and such models are widely believed not to admit exact solutions. Motivated by the search for exact solvability rooted in symmetry, we revisit this problem by uncovering the symmetry structure of general spin–boson Hamiltonians and demonstrating the existence and explicit construction of their exact solutions. 

{\it Symmetry of the models.---} One well-known symmetry of Hamiltonian (\ref{eq:HMSpinBos}) is its conserved parity, arising from the joint parity of the spin and bosons. This discrete symmetry $\mathbb{Z}_2$ plays a central role in organizing the spectrum and constraining the dynamics of the system. The total parity operator is usually defined by $\Pi=\sigma_z\mathcal{P}$, where $\sigma_z$ and $\mathcal{P}=(-1)^{\sum_l^Na_l^{\dagger}a_l}$ are the parity of spin and bosons, respectively. However, a commonly overlooked symmetry is the time-reversal symmetry, usually expressed by the anti-unitary operator $\mathcal{T}$. The function of the operator is specified by
$\mathcal{T}a_j\mathcal{T}^{-1}=a_j$, $\mathcal{T}a_j^{\dagger}\mathcal{T}^{-1}=a_j^{\dagger}$, $\mathcal{T}\sigma_x\mathcal{T}^{-1}=\sigma_x$, and $\mathcal{T}\sigma_z\mathcal{T}^{-1}=\sigma_z$. Therefore, the total symmetry of Hamiltonian (\ref{eq:HMSpinBos}) can be characterized by $\mathcal{T}$, $\Pi$, and $\mathcal{T}\Pi$. 

Examining the symmetry more closely allows us to construct a general procedure for diagonalizing the Hamiltonian in spin space. A rotation about the spin-$X$ axis generated by the bosonic number operator yields the basic transformation rules $e^{i\theta\sigma_x a^\dagger a}ae^{-i\theta \sigma_x a^\dagger a}=e^{-i\theta\sigma_x}a$, and
$e^{i\theta\sigma_x a^\dagger a}a^{\dagger}e^{-i\theta \sigma_x a^\dagger a}=e^{i\theta\sigma_x} a^{\dagger}$ (see Supplementary I for the proof). These relations allow us to introduce or remove exponential Pauli-$X$ factors coupled to the bosonic operators. The normal parity operator can, in fact, be viewed as a special case of such rotations: it rotates the spin and each bosonic mode halfway toward the identity, leaving the interaction term invariant as a whole and thereby preserving the symmetry of the total Hamiltonian. In contrast, our transformations implement arbitrary spin rotations generated by the bosonic number operators, equivalently acting as bosonic rotations conditioned on the spin state. This leads to a richer set of transformation patterns and alters the symmetry structure of the Hamiltonian. In particular, such variations can render the action of the operator $\mathcal{T}$ nontrivial.

\begin{figure}[htbp]
\centering
\includegraphics[width=3.4in]{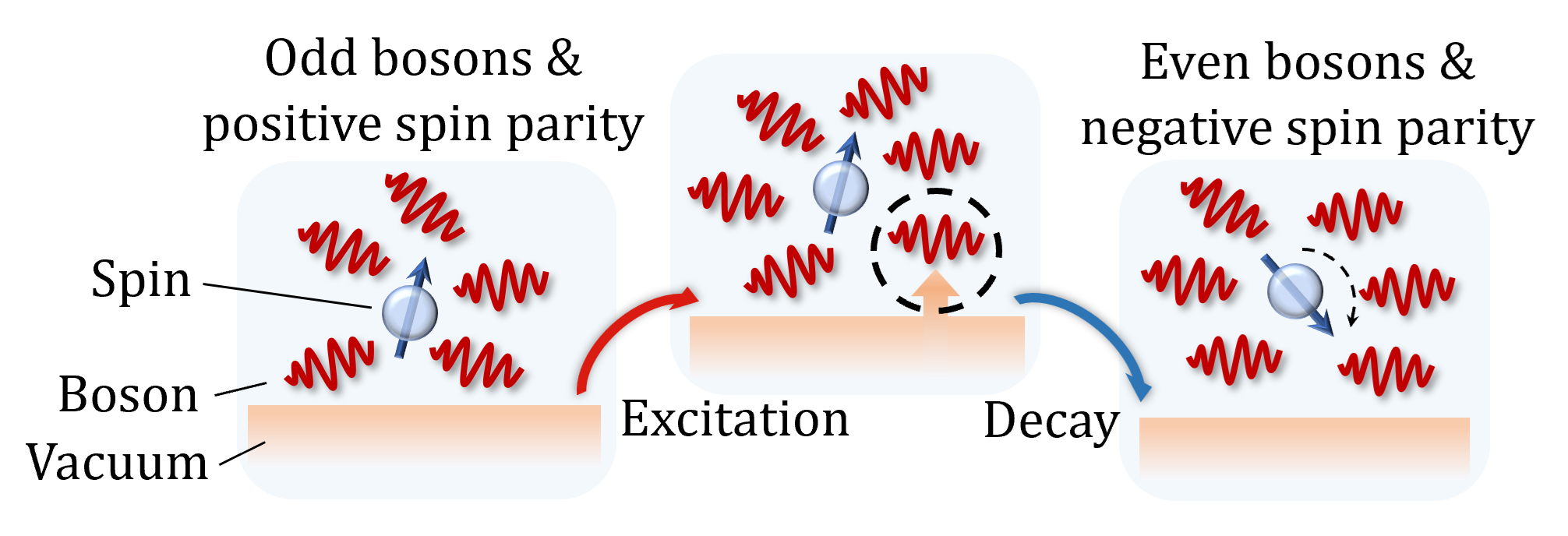}
\caption{A schematic illustration of how would the state evolves under Hamiltonian \eqref{eq:diago}. The energy of the system depends on the parities of both the spin and the bosonic modes. If the number of bosons changes, for example, when the system is excited from an odd-boson-number (lower-energy) state to an even-boson-number (higher-energy) state, it could later relax by flipping the spin parity from positive to negative.}\label{Fig:F1}
\end{figure}

Applying the transformations with $\theta=\pi/2$ and extending the transformation to the multimode case, one obtains the following Hamiltonian 
\begin{equation}\label{eq:diago}
U^{\dagger}H_{M}U  =\sum_{j=1}^{N}\omega_{j}a_{j}^{\dagger}a_{j}-i\sum_{k=1}^{N}g_k(a_{k} ^{\dagger}-a_{k})+\Delta\Pi,
\end{equation}
where $U=e^{i(\pi/2)\sigma_x\sum_{l=1}^Na_{l}^{\dagger}a_{l}}$. It can be seen that the Hamiltonian \eqref{eq:diago} is manifestly diagonal in the spin sector. The symmetry structure of the  Hamiltonian is also modified by this transformation. While the spin-parity symmetry is preserved, the bosonic sector retains only the combined parity–time-reversal symmetry. Consequently, the transformed Hamiltonian is characterized by the symmetry operators $\sigma_z$, $\mathcal{PT}$, and $\sigma_z\mathcal{PT}$. The resulting model describes a qualitatively different spin–boson system. In this system, the crosstalk between the spin and the bosonic modes becomes explicitly parity dependent, as manifested in the third term of right-hand-side of Eq. \eqref{eq:diago}. The second term, proportional to the weighted sum of bosonic momenta, induces flips of the bosonic parity and further affects the effective parity of the spin through the coupling. Although no physical system of this form has yet been recognized, the transformed Hamiltonian remains closely connected to the generalized spin–boson model, a fundamental framework for light–matter interactions as presented in the introduction. The two Hamiltonians share the same spectrum, and their solvability properties are also directly linked. This correspondence provides an interesting insight into each model through the other. A schematic illustration of the state evolution under Hamiltonian \eqref{eq:diago} is shown in Fig. \ref{Fig:F1}.

An equivalent form of Hamiltonian \eqref{eq:diago} is obtained by applying a $\pi/2$ rotation of the bosonic modes. Under the transformation specified by the operator $U_B=e^{-i(\pi/2)\sum_{l=1}^Na_{l}^{\dagger}a_{l}}$, the interaction term in 
Hamiltonian \eqref{eq:diago} becomes $\sum_{k=1}^{N}g_k(a_{k} ^{\dagger}+a_{k})$, while the other terms remain unchanged. Importantly, the single-mode case ($N=1$) of this resulting Hamiltonian coincides with the form previously employed to derive the exact solution of the quantum Rabi model \cite{Braak2011}.

{\it One extension.---} The model described by Hamiltonian \eqref{eq:HMSpinBos} can be extended to
\begin{equation}\label{eq:HMSpinBos-E} H_{MS}=\sum_{j=1}^{N}\omega_{j}a_{j}^{\dagger}a_{j}+\sigma_x\sum_{k=1}^{N}g_k(a_{k} ^{\dagger2}+a^2_{k})+\Delta\sigma_z,
\end{equation} 
which represents a single spin interacting with a multimode squeezed bosonic field. The single-mode case of this model, known as the two-photon Rabi model, has been widely explored in various physical settings \cite{Gerry1988}, including quantum dots embedded in a QED microcavity \cite{Valle2010} and trapped-ion platforms \cite{Felicetti2015}. Its spectral properties have attracted considerable attention \cite{Chen2012,Travenec2012,Maciejewski}, and several isolated exact solutions have been identified \cite{Zhang2014,Zhang2015}. However, the symmetry and solvability of the multi-mode generalization have rarely been explored.

By squaring the transformations in the previous section,  the transformation rules $e^{i\theta\sigma_x a^\dagger a}aae^{-i\theta \sigma_x a^\dagger a}=e^{-i2\theta\sigma_x} aa$, and $e^{i\theta\sigma_x a^\dagger a}a^{\dagger}a^{\dagger}e^{-i\theta \sigma_x a^\dagger a}=e^{i2\theta\sigma_x} a^{\dagger}a^{\dagger}$ can be obtained, which can be further applied to decouple the spin from the squeezed bosonic field. Similarly, applying the squared transformations with $\theta=\pi/4$ and extending the transformation to the multi-mode case, the transformed Hamiltonian can be given by 
\begin{equation}\label{eq:diago-2}
\tilde{U}^{\dagger}H_{MS}\tilde{U}  =\sum_{j=1}^{N}\omega_{j}a_{j}^{\dagger}a_{j}-i\sum_{k=1}^{N}g_k(a_{k}^{\dagger2}-a_{k}^2)+\Delta\Pi_{q},
\end{equation}
where $U=e^{i(\pi/4)\sigma_x\sum_{l=1}^Na_{l}^{\dagger}a_{l}}$ and
\begin{equation*}
\Pi_{q}=\sigma_z\cos{\left(\frac{\pi}{2} \sum_{l=1}^Na_{l}^{\dagger}a_{l}\right)}+\sigma_y\sin{\left(\frac{\pi}{2} \sum_{l=1}^Na_{l}^{\dagger}a_{l}\right)}.   
\end{equation*}
In fact, the operator $\Pi_q$ encodes the $\mathbb{Z}_4$ parity of the system, representing a natural extension of the $\mathbb{Z}_2$ symmetry inherent in Hamiltonian \eqref{eq:HMSpinBos}. It plays the same role as the third term in Hamiltonian \eqref{eq:diago}, and constitutes a key ingredient underlying the solvability of the model described by $H_{MS}$.

The discrete $\mathbb{Z}_4$ symmetry of $H_{MS}$ is specifically generated by $\sigma_z\sqrt{\mathcal{P}}$ with $\sqrt{\mathcal{P}}=e^{i(\pi/2)\sum_{l=1}^N a_l^{\dagger}a_l}$. This implies that the Hilbert space decomposes into four invariant subspaces labeled by the eigenvalues of $\sqrt{\mathcal{P}}$. As revealed by Hamiltonian \eqref{eq:diago-2}, the spin–boson interaction decouples, and the spin contribution is transformed into a symmetry-induced operator. This reformulation produces an explicit decoupling structure for the entire Hamiltonian. Because the operators $a_{k}^{\dagger2}$ and $a_{k}^2$ change the boson number by exactly two, it remains even or odd so that the bosonic parity is strictly preserved. Hence, the third term can be treated independently within the even and odd boson-number sectors, and the full Hilbert space naturally splits into two disjoint subspaces, $\mathcal{H}=\mathcal{H}_e\oplus\mathcal{H}_o$, corresponding to even and odd total bosonic parity. The effective Hamiltonians that govern the two subspaces are  
\begin{equation}
H^{\pm}_{MS}=\sum_{j=1}^{N}\omega_{j}a_{j}^{\dagger}a_{j}-i\sum_{k=1}^{N}g_k(a_{k}^{\dagger2}-a_{k}^2)+\Delta\Pi^{\pm}_{q},
\end{equation}
where $\Pi^{+}_{q}=\sigma_z\sqrt{\mathcal{P}}$ for even boson-number states and $\Pi^{-}_{q}=-i\sigma_y\sqrt{\mathcal{P}}$ for odd boson-number states. Within each invariant subspace, $H^{\pm}_{MS}$ can be further block-diagonalized in the spin basis, resulting in a four-block decomposition of the full Hamiltonian. This decomposition directly reflects the underlying $\mathbb{Z}_4$ symmetry of the system. 

{\it Exact solutions.---} The exact solution of $H_M$ can be obtained more transparently after diagonalizing it in the spin sector. A convenient and powerful method for achieving this is to work in the Bargmann representation \cite{Bargmann1961,Bargmann1962}, an analytic function space that is isomorphic to the Hilbert space of bosonic modes. The key correspondence between bosonic operators and analytic functions in the Bargmann space is $a^{\dagger}\leftrightarrow z$, and $a\leftrightarrow \partial_z$. Furthermore, the correspondence between Bargmann-space wavefunctions and Fock states is given by $|n\rangle\leftrightarrow z^n/\sqrt{n!}$. With this treatment, the Schr{\"o}dinger eigenvalue equation for the original bosonic Hamiltonian is transformed into an eigenvalue problem for analytic functions of complex variables. To treat systems with multiple bosonic modes, one introduces the multi-variable Bargmann space \cite{Vukics2018,Chabaud2022}, in which each mode is represented by its own complex coordinate and the full Hilbert space is realized as analytic functions on $\mathbb{C}^N$. For clarity and ease of comparison with known results, we present the solution in the formalism of Ref.~\cite{Braak2011}, although other equivalent formulations are equally valid.

Using the operator $U$ and $\pi/2$ rotation of the bosonic modes, the following Hamiltonian can be given, 
\begin{equation}\label{eq:HD}
H_D=\sum_{j=1}^{N}\omega_{j}a_{j}^{\dagger}a_{j}+\sum_{k=1}^{N}g_k(a_{k}^{\dagger}+a_{k})+\Delta\Pi.
\end{equation}
In the spin sector, the Hamiltonian $H_D$ becomes block-diagonal, $H_D=\mathrm{diag}(\tilde{H}^{+}_D,\tilde{H}^{-}_D)$. Using multi-varibale Bargammn representation, these operators take the form 
\begin{equation}\label{eq:HD-compo}
\begin{split}
\tilde{H}_D^{\pm}&=\sum_{j=1}^{N}\omega_{j}z_{j}\partial_{z_j}+\sum_{k=1}^{N}g_k(z_{k}+\partial_{z_k})\\
&\pm\Delta(-1)^{\sum_{l=1}^{N}z_{l}\partial_{z_l}}\\
=&(\boldsymbol{\omega}\circ\boldsymbol{z})
\cdot\boldsymbol{\partial_{z}}+\boldsymbol{g}\cdot(\boldsymbol{z}+\boldsymbol{\partial_{z}})\pm\Delta T,    
\end{split}
\end{equation}
where $\circ$ denotes the Hadamard product \cite{Styan1973}. 
$\boldsymbol{\omega}$, $\boldsymbol{g}$, $\boldsymbol{z}$, and $\boldsymbol{\partial_z}$ denote the corresponding vectors of parameters, variables, and partial derivatives. The operator $T$ represents the reflection of all mode variables, defined by $Tf(\boldsymbol{z})=f(\boldsymbol{-z})$ (see Supplementary II for the proof).

By solving the Hamiltonians $H_D^{\pm}$, one obtains the corresponding energy spectra of the two sectors. 
The associated Schr{\"o}dinger equation reads
\begin{equation}
(\boldsymbol{\omega}\circ\boldsymbol{z}+\boldsymbol{g})
\cdot\boldsymbol{\partial_{z}}\psi(\boldsymbol{z})+(\boldsymbol{g}\cdot\boldsymbol{z}-E)\psi(\boldsymbol{z})\pm\Delta\psi(\boldsymbol{-z})=0.
\end{equation}
The equation can be solved using the $G$-function method \cite{Braak2011}. Specifically, the $G$-function associated with this system is given by
\begin{equation}\label{eq:Gfunc}
G_N^{\pm}(X)=\sum_{\boldsymbol{n}\in\mathbb{N}^N}A_{\boldsymbol{n}}\left(1\mp\frac{\Delta}{X-\boldsymbol{\omega}\cdot\boldsymbol{n}}\right)\boldsymbol{g}^{\boldsymbol{n}}, 
\end{equation}
where $\boldsymbol{n}$ denotes an $N$-component multi-index vecotr of non-negative integers, and the summation over $\boldsymbol{n}$ indicates that each component runs independently from 0 to $\infty$. The multi-index monomial $\boldsymbol{g}^{\boldsymbol{n}}$ is defined by $g_1^{n_1}g_2^{n_2}\ldots g_N^{n_N}$. The coefficients $A_{\boldsymbol{n}}$ is given by
\begin{equation}\label{eq:An}
\begin{split}
\sum_{j=1}^Ng_jn_jA_{\boldsymbol{n-1}+\boldsymbol{e}_j}=f_{\boldsymbol{n-1}}A_{\boldsymbol{n-1}}-\sum_{j=1}^Ng_jA_{\boldsymbol{n-1}-\boldsymbol{e}_j},
\end{split}
\end{equation}
Here, $\boldsymbol{e}_j$ denotes an $N$-component unit vector whose $j$-th entry is 1 and all other entries are 0. $\boldsymbol{1}$ denotes an $N$-component vector with all entries being 1. $X$ is defined by $E+\sum_jg_j^2/\omega_j$. $f_{\boldsymbol{n}}(X)$ is given by
\begin{equation}
    f_{\boldsymbol{n}}(X)=2\sum_{j=1}^N\frac{g_j^2}{\omega_j}+\frac{1}{2}\left(\boldsymbol{n}\cdot\boldsymbol{\omega}-X+\frac{\Delta^2}{X-\boldsymbol{n}\cdot\boldsymbol{\omega}}\right).
\end{equation}
The spectrum of the Hamiltonian (\ref{eq:HMSpinBos}) is determined by the zeros of the functions $G^{\pm}_N(X)$. In particular, the zeros of $G^{+}_N(X)$ correspond to the spectrum of the positive-parity sector, while those of $G^{-}_N(X)$ correspond to the negative-parity sector. The initial condition for Eq.~(\ref{eq:An}) can be set by $A_{\boldsymbol{0}}=1$, where $\boldsymbol{0}$ is the zero vector. To obtain each $A_{\boldsymbol{n}}$, one needs the analytical requirements of $\psi(\boldsymbol{z})$ and permutation symmetry of the bosons. From Hamiltonian (\ref{eq:HMSpinBos}), one can observe that the spectrum of the system remains invariant under the simultaneous exchanges $g_j \leftrightarrow g_k$ and $\omega_j \leftrightarrow \omega_k$. This property effectively reduces the number of degrees of freedom, thereby ensuring the solvability of Eq.~(\ref{eq:An}) and ultimately yielding an exact solution of the Hamiltonian. Further details are provided in Supplementary III.

The above results are consistent with previously known solutions. As a primary example, in the single-mode limit, one can verify that $G_N^{\pm}(X)$ reduces to the standard Rabi model $G$-function when $N=1$. Furthermore, as a numerical verification, the two-mode case ($N=2$) is evaluated. The plot of $G_2^{\pm}$ and the landscape of the first energy level is shown in Figs. \ref{Fig:F2} and \ref{Fig:F3}, respectively. The analytical details are given in Supplementary IV.

\begin{figure}[htbp]
\centering
\includegraphics[width=3.41in]{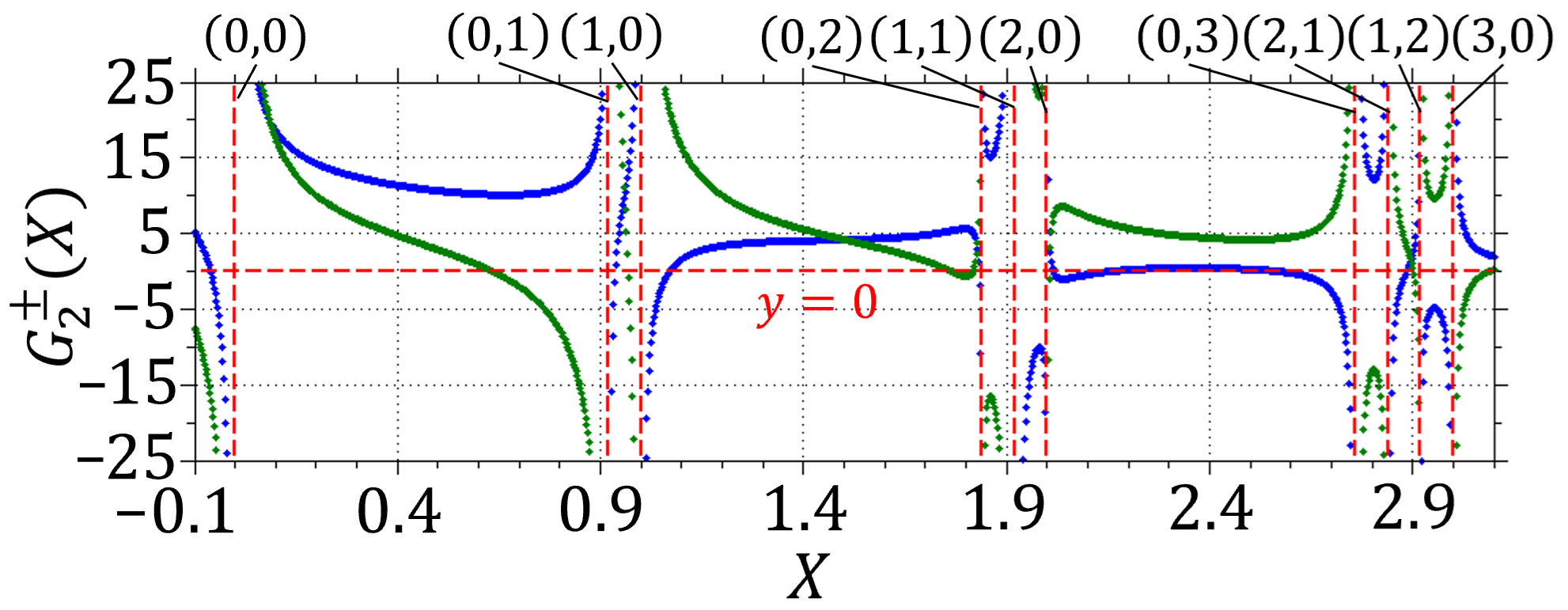}
\caption{The behaviour of $G_2^{+}$ (dotted blue curve) and $G_2^{-}$ (dotted Green curve) when $X\in[-0.1,3.1]$. The parameters are $\omega_1=1,~w_2=0.92$, $g_1=0.7,~g_2=0.78$, and $\Delta=0.68$. The vertical dashed rel lines marks the poles. They are given by $X=n_1\omega_1+n_2\omega_2$, and marked by $(n_1,n_2)$ in the upper place. The line $y=0$ is also marked out.}\label{Fig:F2}
\end{figure}As shown in Fig.~\ref{Fig:F2}, the analytic functions $G_2^{\pm}(X)$ exhibit simple poles at $X = 0\omega_1+0\omega_2,~1\omega_1+0\omega_2,~ 0\omega_1+1\omega_2,\ldots$, which correspond to the bosonic energy levels in the absence of coupling to the spin. The zeros of $G_2^{\pm}(X)$ lie in the vicinity of these poles, consistent with the behavior of the single-mode $G$-function \cite{Braak2011}.
 
Figure~\ref{Fig:F3} displays the landscapes of the lowest energy level as functions of the couplings $g_1$ and $g_2$, obtained from the zeros of $G_2^{\pm}$ under different parameter settings. The upper and lower surface in panel (a) correspond to the positive- and negative-parity sectors, respectively. The overall trends of these landscapes are consistent with the single-mode behavior, as more clearly illustrated by the curves for $g_2=0$ (Fig.~\ref{Fig:F3}(b)) and $g_1=g_2$ (Fig.~\ref{Fig:F3}(c)). Notably, the landscapes are approximately symmetric about the line $g_1=g_2$. This arises because $\omega_1$ and $\omega_2$ are close in value, so exchanging 
$g_1$ and $g_2$ produces only a minor variation in the energy level, in agreement with the permutation symmetry of the bosonic modes discussed above.

\begin{figure}[htbp]
\centering
\includegraphics[width=3.41in]{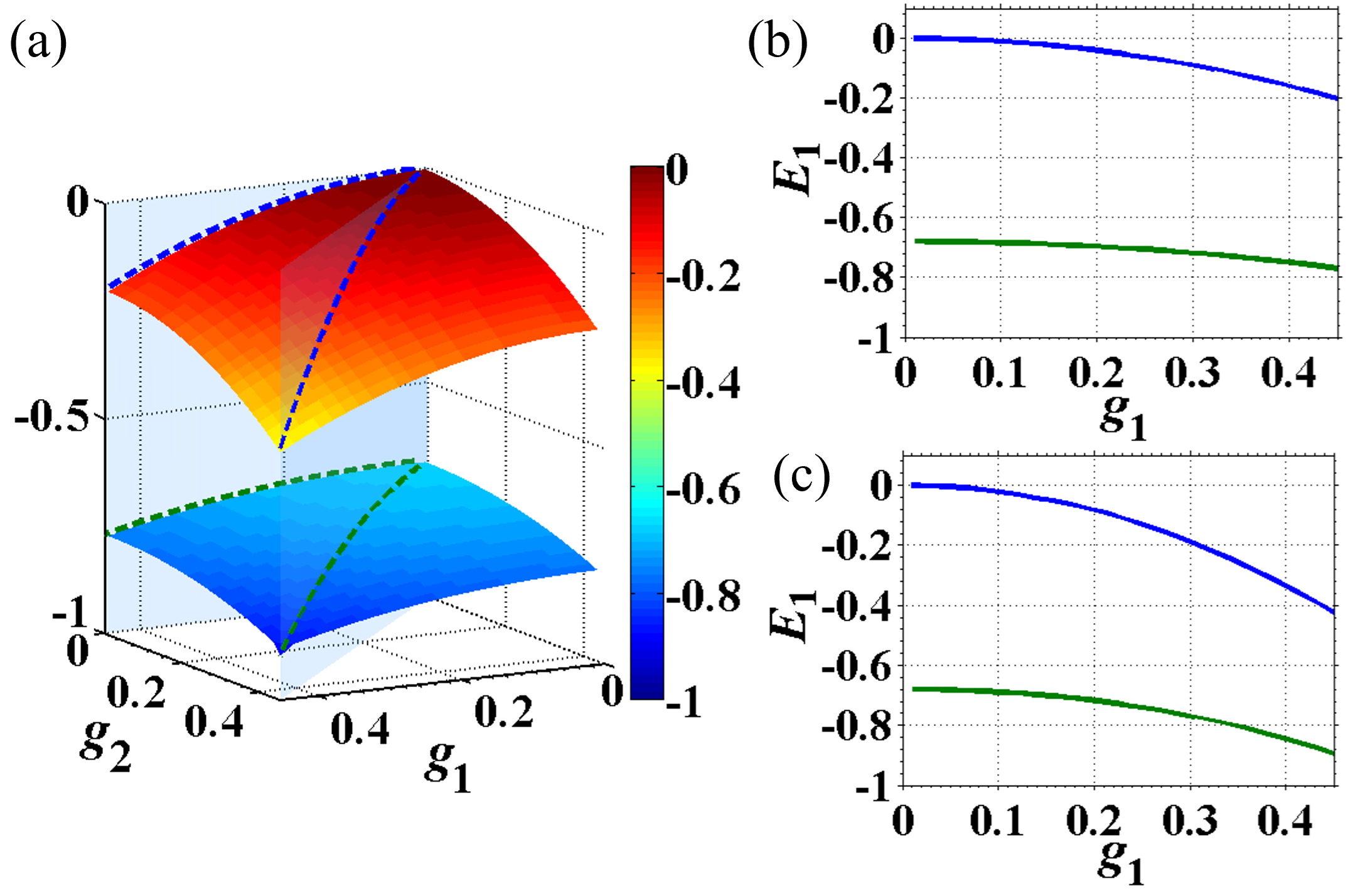}
\caption{Landscape of the first energy level when $g_1,g_2\in[0,0.4]$. (a) The surfaces of the positive and the negative sectors. (b) The plot of the $g_2=0$ edge in (a). (c) The plot of diagonal line ($g_1=g_2$) in (a). The dashed lines in (a) mark out the position for obtains the curves in (b) and (c), colored accordingly. $\omega_1=1$;  $\omega_2=0.92$.}\label{Fig:F3}
\end{figure}

{\it Conclusions.---} In this work, we highlight two fundamental symmetries of generalized spin–boson models that ultimately enable their exact solution. The first is time-reversal symmetry, which lies at the root of the general conventional expressions for diagonalizing the Hamiltonian. The second is the permutation symmetry of the bosonic modes, which underpins the construction of the series expansion underlying the generalized $G$-function.

We further emphasize that the diagonalizations given in Eqs.~(\ref{eq:diago}) and (\ref{eq:diago-2}) reveal a previously unnoticed connection between generalized spin–boson models and a parity-interaction model. Finally, the theoretical framework developed here opens the way for analyzing the intricate energy spectra of generalized spin–boson systems in high-dimensional Hilbert spaces, which we leave for future investigation.

\acknowledgments
This work is funded by MCIN/AEI /10.13039/501100011033 (No. PID2021-126273NB-I00), and by ``ERDF A way of making Europe'', and by the Basque Government through Grant No. IT1470-22. This project has also received support from the Spanish Ministry of Economic Affairs and Digital Transformation through the QUANTUM ENIA project call -Quantum Spain, and by the EU through the Recovery, Transformation and Resilience Plan- NextGenerationEU within the framework of the Digital Spain 2026 Agenda. We also acknowledge the support by the National Natural Science Foundation of China (No.12474355).

\clearpage
\onecolumngrid
\section*{Supplementary of ``Symmetry and Exact Solutions of General Spin-Boson Models''}
\setcounter{equation}{0}
\renewcommand{\theequation}{S\arabic{equation}}
\setcounter{figure}{0}
\renewcommand{\thefigure}{S\arabic{figure}}



\section{I. Pauli-X rotations generated by the bosonic number operator}
According to the Baker–Campbell–Hausdorff (BCH) formula, 
\begin{equation}
\begin{split}
    e^{-B}Ae^B=A+[A,B]+\frac{1}{2!}[[A,B],B]+\frac{1}{3!}[[[A,B],B],B]+\cdots,
\end{split}
\end{equation}
one has 
\begin{equation}\label{eq:trans-1}
    e^{i\theta a^\dagger a}ae^{-i\theta a^\dagger a}=e^{-i\theta}a,~e^{i\theta a^\dagger a}a^{\dagger}e^{-i\theta a^\dagger a}=e^{i\theta}a^{\dagger},
\end{equation}
and 
\begin{equation}\label{eq:trans-2}
    \begin{split}
    e^{\alpha^*a-\alpha a^{\dagger}}ae^{\alpha a^{\dagger}-\alpha^*a}&=a+\alpha,\\
    e^{\alpha^*a-\alpha a^{\dagger}}a^{\dagger}e^{\alpha a^{\dagger}-\alpha^*a}&=a^{\dagger}+\alpha^*.
    \end{split}
\end{equation}
The above calculation is based on the commutator $[a, a^{\dagger}]=1$. Using Eq. (\ref{eq:trans-1}), one has
\begin{equation}\label{eq:trans-3}
    e^{i\theta\sigma_x a^\dagger a}ae^{-i\theta \sigma_xa^\dagger a}=e^{-i\theta\sigma_x}a,
\end{equation}
by simply applying $\theta\rightarrow\theta\sigma_x$. This is valid because $\sigma_x$ and $a$ only give a tensor product. It will not change the BCH expansion form of Eq. (\ref{eq:trans-1}). The complete formula is  
\begin{equation}
   e^{i\theta\sigma_x\otimes a^\dagger a}ae^{-i\theta \sigma_x\otimes a^\dagger a}=e^{-i\theta\sigma_x}\otimes a. 
\end{equation}
Recalls that $\exp(-i\theta\sigma_x)$  equals $I\cos{\theta}+i\sigma_x\sin{\theta}$. Then, set $\theta =\pi/2$, one has
\begin{equation}
    \begin{split}
    e^{i\frac{\pi}{2}\sigma_x a^\dagger a}ae^{-i\frac{\pi}{2} \sigma_xa^\dagger a}&=-i\sigma_xa,\\
    e^{i\frac{\pi}{2}\sigma_x a^\dagger a}a^{\dagger}e^{-i\frac{\pi}{2} \sigma_xa^\dagger a}&=i\sigma_xa^{\dagger}.
    \end{split}
\end{equation}
Therefore, one has
\begin{equation}\label{eq:relation-1}
        e^{-i\frac{\pi}{2}\sigma_x a^\dagger a}\sigma_x(a^{\dagger}+a)e^{i\frac{\pi}{2} \sigma_xa^\dagger a}=-i(a^{\dagger}-a).
\end{equation}
Then, for the Pauli-Z operator, one has
\begin{equation}\label{eq:trans-4}
\begin{split}
     e^{-i\frac{\pi}{2}\sigma_x}\sigma_ze^{i\frac{\pi}{2}\sigma_x}&=\left[\cos\left(-\frac{\pi}{2}\right)I+i\sin\left(-\frac{\pi}{2}\right)\sigma_x\right]\sigma_z\left[\cos\left(\frac{\pi}{2}\right)I+i\sin\left(\frac{\pi}{2}\right)\sigma_x\right]\\
     &=(\cos\pi)\sigma_z+i\sin\left(\frac{\pi}{2}\right)\cos\left(\frac{\pi}{2}\right)[\sigma_z,\sigma_x]\\
     &=(\cos\pi)\sigma_z-(\sin{\pi})\sigma_y\\
     &=-\sigma_z.
\end{split}
\end{equation}
By adding $a^{\dagger}a$, one has
\begin{equation}\label{eq:relation-2}
    \begin{split}
        &e^{-i\frac{\pi}{2}\sigma_x\otimes a^{\dagger}a}(\sigma_z\otimes I)e^{i\frac{\pi}{2}\sigma_x\otimes a^{\dagger}a}\\
        =&\sigma_z\otimes\cos(\pi a^{\dagger}a)-\sigma_y\otimes\sin({\pi}a^{\dagger}a)\\
        =&\sigma_z\otimes(-1)^{a^{\dagger}a}.
    \end{split}
\end{equation}
The last equation can be seen by applying $\cos(\pi a^{\dagger}a)$ and $(-1)^{a^{\dagger}a}$ to a Fock state $|n\rangle$.

Finally, using Eq. (\ref{eq:relation-1}) and Eq. (\ref{eq:relation-2}), one can define the transformation operator $\exp[-i(\pi/2)\sigma_x\sum_la_{l}^{\dagger}a_{l}]$. Because $[a_j,a_k]=0$ and $[a^{\dagger}_j,a_k]=0$ if $j\ne k$, one can apply the transformation to obtain the diagonal form of the spin-boson Hamiltonian in the main text. 

The transformation relation for the squeezed bosonic field interaction Hamiltonian can be given by squaring the Eq. (\ref{eq:trans-3}). For example, squaring the transformation on $a$, the left-hand side reads
\begin{equation}
    \begin{split}
    &(e^{i\theta\sigma_x\otimes a^\dagger a}ae^{-i\theta\sigma_x\otimes a^\dagger a})(e^{i\theta\sigma_x \otimes a^\dagger\sigma_x a}ae^{-i\theta\sigma_x\otimes a^\dagger a})\\
    &=e^{i\theta\sigma_x\otimes a^\dagger a}aae^{-i\theta\sigma_x\otimes a^\dagger a},
    \end{split}
\end{equation}
and the right-hand side reads
\begin{equation}
(e^{i\theta\sigma_x}\otimes a)(e^{i\theta\sigma_x}\otimes a)=e^{i2\theta\sigma_x}\otimes(aa).
\end{equation}
The transformation on the Pauli operators is similar to Eq. (\ref{eq:relation-2}).

\section{II. The reflection operator in Bargmann space.}
The mapping from boson operators to the Bargmann space is given by
\begin{equation}
    a_j^{\dagger}\rightarrow z_j,~a_j\rightarrow\frac{\partial}{\partial z_j}.
\end{equation}
As shown in the main text, the map transforms the last term of the Spin-Boson Hamiltonian  into
\begin{equation}
    \Delta\sigma_z(-1)^{(z_1\frac{\partial}{\partial z_1}+z_2\frac{\partial}{\partial z_2}+\cdots)}.
\end{equation}
Next, we use the two-term form to show the case, and the general case is the same. Notice that $e^{i\pi}=-1$. Therefore, in the Bargmann space, the operator should be
\begin{equation}
    (e^{i\pi})^{(z_1\frac{\partial}{\partial z_1}+z_2\frac{\partial}{\partial z_2})}=e^{i\pi(z_1\frac{\partial}{\partial z_1}+z_2\frac{\partial}{\partial z_2})}.
\end{equation}
Because the Taylor expansion of an exponential function shows good convergence, one can use the expansion for computation. For the factor $z_1^mz_2^n/\sqrt{m!n!}$ (corresponding to $|m\rangle|n\rangle$ in the Fork space), it is given by
\begin{equation}
\begin{split}
    &(-1)^{(z_1\frac{\partial}{\partial z_1}+z_2\frac{\partial}{\partial z_2})}\frac{z_1^mz_2^n}{\sqrt{m!n!}}\\
    =&e^{i\pi(z_1\frac{\partial}{\partial z_1}+z_2\frac{\partial}{\partial z_2})}\frac{z_1^mz_2^n}{\sqrt{m!n!}}=e^{i\pi z_1\frac{\partial}{\partial z_1}}e^{i\pi z_2\frac{\partial}{\partial z_2}}\frac{z_1^mz_2^n}{\sqrt{m!n!}}\\
    =&\left[\sum_{a=0}^{\infty}\frac{1}{a!}(i\pi z_1\frac{\partial}{\partial z_1})^a\frac{z_1^m}{\sqrt{m!}}\right]\left[\sum_{b=0}^{\infty}\frac{1}{b!}(i\pi z_2\frac{\partial}{\partial z_2})^b\frac{z_2^n}{\sqrt{n!}}\right].
\end{split}
\end{equation}
Using a simple relation, the above analysis can be simplified. Notice that
\begin{equation}
    (z\frac{\partial}{\partial z})^k z^n=n^kz^n.
\end{equation}
This is because
\begin{equation}
\begin{split}
      &(z\frac{\partial}{\partial z})\cdots(z\frac{\partial}{\partial z})(z\frac{\partial}{\partial z})z^n\\
    &=(z\frac{\partial}{\partial z})\cdots(z\frac{\partial}{\partial z})nz^n=(z\frac{\partial}{\partial z})\cdots n^2z^n\\
    &=\cdots=n^kz^n.
\end{split}
\end{equation}
Therefore, more generally, for an element in the Bargmann space $f(z)$, one has
\begin{equation}
\begin{split}
    &e^{\lambda z\frac{d}{dz}}f(z)=\sum_{k=0}^{\infty}\frac{\lambda^k}{k!}(z\frac{d}{dz})^k(\sum_{n=0}^{\infty}c_nz^n)\\
    =&\sum_{n=0}^{\infty}c_n\left[\sum_{k=0}^{\infty}\frac{\lambda^k}{k!}(z\frac{d}{dz})^kz^n\right]=\sum_{n=0}^{\infty}c_n\left[\sum_{k=0}^{\infty}\frac{(\lambda n)^k}{k!}z^n\right]\\
    =&\sum_{n=0}^{\infty}c_n\left[\sum_{k=0}^{\infty}\frac{(\lambda n)^k}{k!}\right]z^n=\sum_{n=0}^{\infty}c_ne^{n\lambda}z^n\\
    =&\sum_{n=0}^{\infty}c_n(e^{\lambda}z)^n=f(e^\lambda z).
\end{split}
\end{equation}
Let $\lambda=i\pi$, and one can go back to the previous case. Therefore,
\begin{equation}
\begin{split}
    (-1)^{(z_1\frac{\partial}{\partial z_1}+z_2\frac{\partial}{\partial z_2})}\frac{z_1^mz_2^n}{\sqrt{m!n!}}=(-1)^{m+n}\frac{z_1^mz_2^n}{\sqrt{m!n!}}=\frac{(-z_1)^m(-z_2)^n}{\sqrt{m!n!}},
\end{split}
\end{equation}
and, moreover,
\begin{equation}
    (-1)^{(z_1\frac{\partial}{\partial z_1}+z_2\frac{\partial}{\partial z_2})}f(z_1,z_2)=f(-z_1,-z_2).
\end{equation}
This means that $(-1)^{(z_1\frac{\partial}{\partial z_1}+z_2\frac{\partial}{\partial z_2})}$ actually equals the $\hat{T}$ operator in the following part, or Braak's method. As mentioned, the gneral multi-mode case is the same.

\section{III. Derivation of solutions to General Spin-Boson Hamiltonians}
We start from Schr{\"o}dinger the equation,  \begin{equation}\label{eq:origin}
(\boldsymbol{\omega}\circ\boldsymbol{z}+\boldsymbol{g})
\cdot\boldsymbol{\partial_{z}}\psi(\boldsymbol{z})+(\boldsymbol{g}\cdot\boldsymbol{z}-E)\psi(\boldsymbol{z})\pm\Delta\psi(\boldsymbol{-z})=0,
\end{equation}
where 
\begin{equation}
\begin{split}
 \boldsymbol{\omega}&=\omega_1,\ldots,\omega_N),\\
 \boldsymbol{z}&=(z_1,\ldots,z_N),\\
 \boldsymbol{g}&=(g_1,\ldots,g_N),\\ \partial_{\boldsymbol{z}}&=(\frac{\partial}{\partial z_1},\ldots,\frac{\partial}{\partial z_N}),\\ \boldsymbol{\omega}\circ\boldsymbol{z}&=(\omega_1z_1,\ldots,\omega_Nz_N)
\end{split}
\end{equation}

We take the positive case as an example. By using the method presented by D. Braak, the solution to the equation is encoded by two coupled equations
\begin{equation}
\begin{split}
(\boldsymbol{\omega}\circ\boldsymbol{z}+\boldsymbol{g})
\cdot\boldsymbol{\partial_{z}}\psi(\boldsymbol{z})+(\boldsymbol{g}\cdot\boldsymbol{z}-E)\psi(\boldsymbol{z})+\Delta\phi(\boldsymbol{z})=0,\\
(\boldsymbol{\omega}\circ\boldsymbol{z}-\boldsymbol{g})
\cdot\boldsymbol{\partial_{z}}\phi(\boldsymbol{z})-(\boldsymbol{g}\cdot\boldsymbol{z}+E)\phi(\boldsymbol{z})+\Delta\psi(\boldsymbol{z})=0,
\end{split}
\end{equation}
under the condition that $\psi(-\boldsymbol{z})=\phi(\boldsymbol{z})$. Notice that the second equation is obtained by performing the reflection operator $T$ in the second equation. 

Firstly, shift the parameter by $\boldsymbol{y}=\boldsymbol{z}+\boldsymbol{g}\circ\boldsymbol{\omega^{-1}}$, where $\boldsymbol{\omega^{-1}}=(1/\omega_1,\ldots,1/\omega_N)$. Then, the equation changes to 
\begin{equation}
\begin{split}
(\boldsymbol{\omega}\circ\boldsymbol{y})
\cdot\boldsymbol{\partial_{y}}\psi(\boldsymbol{y})+[\boldsymbol{g}\cdot&\boldsymbol{y}-(E
+\boldsymbol{g}\cdot(\boldsymbol{g}\circ\boldsymbol{\omega^{-1}})]\psi(\boldsymbol{y})+\Delta\phi(\boldsymbol{y})=0,\\
(\boldsymbol{\omega}\circ\boldsymbol{y}-2\boldsymbol{g})
\cdot\boldsymbol{\partial_{y}}\phi(\boldsymbol{y})-[&\boldsymbol{g}\cdot\boldsymbol{y}+(E-\boldsymbol{g}\cdot(\boldsymbol{g}\circ\boldsymbol{\omega^{-1}})]\phi(\boldsymbol{y})+\Delta\psi(\boldsymbol{y})=0,
\end{split}
\end{equation}
Notice that $\boldsymbol{\partial}_{\boldsymbol{z}}=\boldsymbol{\partial}_{\boldsymbol{y}}$ and $\boldsymbol{\omega}\circ(\boldsymbol{g}\circ\boldsymbol{\omega^{-1}})=\boldsymbol{g}$. 

Secondly, assume that $\psi$ and $\phi$ have exponential scale factors, which guarantees the convergence of the solutions at infinite points. Hence, set 
\begin{equation*}
\begin{split}
\psi(\boldsymbol{y})&=e^{-\boldsymbol{g}\circ\boldsymbol{\omega^{-1}}\cdot(\boldsymbol{y}-\boldsymbol{g}\circ\boldsymbol{\omega^{-1}})}\bar\psi(\boldsymbol{y}),\\    
\phi(\boldsymbol{y})&=e^{-\boldsymbol{g}\circ\boldsymbol{\omega^{-1}}\cdot(\boldsymbol{y}-\boldsymbol{g}\circ\boldsymbol{\omega^{-1}})}\bar\phi(\boldsymbol{y}).
\end{split}
\end{equation*}
Then, the set of equations changes to
\begin{equation}\label{eq:MMeqset}
\begin{split}
(\boldsymbol{\omega}\circ\boldsymbol{y})
\cdot\boldsymbol{\partial_{y}}\bar\psi(\boldsymbol{y})&-X\bar\psi(\boldsymbol{y})+\Delta\bar\phi(\boldsymbol{y})=0,\\
(\boldsymbol{\omega}\circ\boldsymbol{y}-2\boldsymbol{g})
\cdot\boldsymbol{\partial_{y}}\bar\phi(\boldsymbol{y})-[2\boldsymbol{g}\cdot\boldsymbol{y}&+X
-4\boldsymbol{g}\cdot(\boldsymbol{g}\circ\boldsymbol{\omega^{-1}})]\bar\phi(\boldsymbol{y})+\Delta\bar\psi(\boldsymbol{y})=0,
\end{split}
\end{equation}
where $X$ is defined by $E+\boldsymbol{g}\cdot(\boldsymbol{g}\circ\boldsymbol{\omega^{-1}})$. This is based on the partial derivative
\begin{equation*}
\begin{split}
&\boldsymbol{\partial_{y}}[e^{-\boldsymbol{g}\circ\boldsymbol{\omega^{-1}}\cdot(\boldsymbol{y}-\boldsymbol{g}\circ\boldsymbol{\omega^{-1}})}\bar\psi(\boldsymbol{y})\\
=&-\boldsymbol{g}\circ\boldsymbol{\omega^{-1}}[e^{-\boldsymbol{g}\circ\boldsymbol{\omega^{-1}}\cdot(\boldsymbol{y}-\boldsymbol{g}\circ\boldsymbol{\omega^{-1}})}\bar\psi(\boldsymbol{y})]+[e^{-\boldsymbol{g}\circ\boldsymbol{\omega^{-1}}\cdot(\boldsymbol{y}-\boldsymbol{g}\circ\boldsymbol{\omega^{-1}})}\boldsymbol{\partial_{y}}\bar\psi(\boldsymbol{y})],  
\end{split}
\end{equation*}
as well as the one for $\bar\phi$. The exponential factors can be canceled out.

Thirdly, consider the polynomial solutions of $\bar\psi$ and $\bar\phi$. Set
\begin{equation}
\begin{split}
    \bar\psi(\boldsymbol{y})&=\sum_{\boldsymbol{n}\in\mathbb{Z}^N}B'_{\boldsymbol{n}}\boldsymbol{y}^{\boldsymbol{n}}\coloneq\sum_{n_1,\ldots,n_N=-\infty}^{\infty}B'_{n_1,\ldots,n_N}y_1^{n_1}y_2^{n_2}\cdots y_N^{n_N},\\
    \bar\phi(\boldsymbol{y})&=\sum_{\boldsymbol{n}\in\mathbb{Z}^N}B_{\boldsymbol{n}}\boldsymbol{y}^{\boldsymbol{n}}\coloneq\sum_{n_1,\ldots,n_N=-\infty}^{\infty}B_{n_1,\ldots,n_N}y_1^{n_1}y_2^{n_2}\cdots y_N^{n_N},
\end{split}
\end{equation}
where $\boldsymbol{n}=(n_1,\ldots,n_N)$. Substitute the solutions to the first of Eqs. (\ref{eq:MMeqset}), one has
\begin{equation}
    B'_{\boldsymbol{n}}=B_{\boldsymbol{n}}\frac{\Delta}{X-\boldsymbol{n}\cdot\boldsymbol{\omega}}.
\end{equation}
Using the above relation and the second of Eqs. (\ref{eq:MMeqset}), one obtains the recurrence relation for $B$,
\begin{equation}
\begin{split}
      &g_1(n_1+1)B_{n_1+1,n_2,\ldots}+g_2(n_2+1)B_{n_1,n_2+1,\ldots}+\cdots\\
      =&f_{n_1,n_2,\ldots}B_{n_1,n_2,\ldots}-g_1B_{n_1-1,n_2,\ldots}-g_2B_{n_1,n_2-1,\ldots}-\cdots.
\end{split}
\end{equation}
In a more compact form, one has
\begin{equation}\label{recurrence}
\sum_{j=1}^Ng_j(n_j+1)B_{\boldsymbol{n}+\boldsymbol{e}_j}=f_{\boldsymbol{n}}B_{\boldsymbol{n}}-\sum_{j=1}^Ng_jB_{\boldsymbol{n}-\boldsymbol{e}_j}. 
\end{equation}
$\boldsymbol{e}_j$ denotes an $N$-component vector whose $j$th component is $1$ and all others are zeros. $f_{\boldsymbol{n}}$ is given by
\begin{equation}\label{eq:fun}
    f_{\boldsymbol{n}}(X)=2\boldsymbol{g}\cdot(\boldsymbol{g}\circ\boldsymbol{\omega}^{-1})+\frac{1}{2}\left(\boldsymbol{n}\cdot\boldsymbol{\omega}-X+\frac{\Delta^2}{X-\boldsymbol{n}\cdot\boldsymbol{\omega}}\right).
\end{equation}
Calculating each $B_{\boldsymbol{n}}$ from the recurrence, the formal solutions to Eq. (\ref{eq:MMeqset}) can be given by
\begin{equation}
\begin{split}
  \psi(\boldsymbol{z})&=e^{-(\boldsymbol{g}\circ\boldsymbol{\omega^{-1}})\cdot\boldsymbol{z}}\sum_{\boldsymbol{n}\in\mathbb{Z}^N}\frac{\Delta}{X-\boldsymbol{n}\cdot\boldsymbol{\omega}}B_{\boldsymbol{n}}(\boldsymbol{z}+\boldsymbol{g}\circ\boldsymbol{\omega}^{-1})^{\boldsymbol{n}},\\    
&\phi(\boldsymbol{z})=e^{-(\boldsymbol{g}\circ\boldsymbol{\omega^{-1}})\cdot\boldsymbol{z}}\sum_{\boldsymbol{n}\in\mathbb{Z}^N}B_{\boldsymbol{n}}(\boldsymbol{z}+\boldsymbol{g}\circ\boldsymbol{\omega}^{-1})^{\boldsymbol{n}}.  
\end{split}
\end{equation}
Like what Braak has done in his work, the energy spectrum (or equivalently $X$) can be obtained by using the analytical property of the solutions: (a) in order to avoid singular points at $\boldsymbol{z}=\boldsymbol{0}$, $B_{\boldsymbol{n}\in(\mathbb{Z^{-}})^N}=0$; (b) in order to obtain the solutions to the original Eq. (\ref{eq:origin}) (the positive sector), $\psi(\boldsymbol{z})$ must equal to $\phi(-\boldsymbol{z})$. Hence, the spectrum, or the eigenvalues of Eq. (\ref{eq:origin}), can be given by the zeros of the function $G^{\pm}_N(X;\boldsymbol{z})=\phi(-\boldsymbol{z})-\psi(\boldsymbol{z})$. Notice that different $\boldsymbol{z}$s in the complex-variable space shall not change the spectrum. Therefore, by taking $\boldsymbol{z}=\boldsymbol{0}$, one can employ
\begin{equation}\label{eq:Gfun}
\begin{split}
G_N^{\pm}(X)\coloneq G_N^{\pm}(X;0)=\sum_{\boldsymbol{n}\in\mathbb{N}^N}B_{\boldsymbol{n}}\left(1\mp\frac{\Delta}{X-\boldsymbol{n}\cdot\boldsymbol{\omega}}\right)(\boldsymbol{g}\circ\boldsymbol{\omega}^{-1})^{\boldsymbol{n}},    
\end{split}
\end{equation}
for finding the zeros. Furthermore, the energy spectrum can be obtained by the zeros. Notice that the spectrum for the negative sector is obtained by making the transformation $\Delta\rightarrow-\Delta$.

Next, we discuss the calculation of $B_{\boldsymbol{n}}$. Although $B_{\boldsymbol{n}\in(\mathbb{Z^{-}})^N}=0$ is set due to the analytical property of the solutions, Eq. (\ref{recurrence}) still holds for all $\boldsymbol{n}\in\mathbb{Z}^N$. Then, one can calculate each $B_{\boldsymbol{n}}$ by substituting different integers into the subscripts. There are many ways to accomplish this, and we give one strategy of them as follows.

Define the total bosonic mode number as $p=\sum_{j=1}^Nn_j$. By evaluating $p$ with integers, or equivalently setting $n_j$s, one can obtain series of linear equation sets for $B_{\boldsymbol{n}}$. Notably, Eq. (\ref{recurrence}) changes to 
\begin{equation}\label{recurrence-1}
\sum_{j=1}^Ng_jn_jB_{\boldsymbol{n}-\boldsymbol{1}+\boldsymbol{e}_j}=f_{\boldsymbol{n-1}}B_{\boldsymbol{n-1}}-\sum_{j=1}^Ng_jB_{\boldsymbol{n-1}-\boldsymbol{e}_j}. 
\end{equation} It can be seen that the cases when $p<N$ are trivial. This is because there are no choices of $\boldsymbol{n}$ in those cases such that the components of the vector $\boldsymbol{n-1}+\boldsymbol{e}_j$ are all non-negative. A simple example is the case when $p=N-1$. In such a case, if $\boldsymbol{n}=(N-1,0,0,\ldots,0)$, $\boldsymbol{n-1}+\boldsymbol{e}_j$ is $(N-2,-1,-1,\ldots)$, which has a large number of negative components. As we mentioned, $B_{\boldsymbol{n}}$ with negative subscripts must be zeros. Another example is the case when $\boldsymbol{n}=(1,1,\ldots,1,0)$. Substituting it into Eq. (\ref{recurrence-1}), one obtains the relation $0\cdot B_{\boldsymbol{0}}=0$, which corresponds to a boundary condition. In the main text, we set $B_{\boldsymbol{0}}=1$. The other choices of $\boldsymbol{n}$ under this condition are similar. 

Things go differently when $p\geq N$. For example, when $p=N$, there is only one effective set for $\boldsymbol{n}$, i.e., $\boldsymbol{n}=\boldsymbol{1}$. Then, one has 
$\sum_{j=1}^Ng_jB_{\boldsymbol{e}_j}=f_{\boldsymbol{0}}B_{\boldsymbol{0}}=f_{\boldsymbol{0}}$. This is an under-determined equation for $B_{\boldsymbol{e}_j}$. There are multiple choices, and one can obtain a confirmed one by introducing extra symmetry constraints, such as the permutation symmetry of the spin-boson Hamiltonian. 

The specific description and the affection of the permutation symmetry are discussed as follows.  From the Hamiltonian, it can be observed that 
\begin{equation}
    [g_i\rightleftharpoons g_j,~\omega_i\rightleftharpoons\omega_j]\Leftrightarrow[a_i^{\dagger}\rightleftharpoons a_j^{\dagger},a_i\rightleftharpoons a_j].
\end{equation}
Furthermore, exchanging the $i$th and $j$th bosonic modes (exchanging $a_i^{\dagger}$ and $a_j^{\dagger}$, and simultaneously exchanging $a_i$ and $a_j$) will not affect the spectrum. There are no differences emerging after the transformations, except the name of.the modes. Therefore, the zeros of Eq. (\ref{eq:Gfun}) must not change. Specifically in Eq. (\ref{eq:Gfun}), if one makes the exchange of $g_1$ and $g_2$, along with the exchange of $\omega_1$ and $\omega_2$, it can be obtained that
\begin{equation}
\begin{split}
&\sum_{\boldsymbol{n}\in\mathbb{N}^N}B_{n_1,n_2,\ldots}\left[1\mp\frac{\Delta}{X-(n_1\omega_2+n_2\omega_1+\ldots)}\right]\left(\frac{g_2}{\omega_2}\right)^{n_1}\left(\frac{g_1}{\omega_1}\right)^{n_2}\ldots\\
=&\sum_{\boldsymbol{n}\in\mathbb{N}^N}B_{n_2,n_1,\ldots}\left[1\mp\frac{\Delta}{X-(n_2\omega_2+n_1\omega_1+\ldots)}\right]\left(\frac{g_2}{\omega_2}\right)^{n_2}\left(\frac{g_1}{\omega_1}\right)^{n_1}\ldots.
\end{split}
\end{equation}
The equation holds because one can change the number of the summation index, for example, rename the one with $\omega_1$ by $n_1$, and the one with $\omega_2$ by $n_2$. Hence, the main factors align with each other, and one has
$B_{n_1,n_2,\ldots}=B_{n_1,n_2,\ldots}(g_1\rightleftharpoons g_2,\omega_1\rightleftharpoons\omega_2)$. It means that $B_{n_1,n_2,\ldots}$ solved from Eq. (\ref{recurrence-1}) equals to $B_{n_2,n_1,\ldots}$ solved from the equation after exchanging $g_1$ and $g_2$ as well as $\omega_1$ and $\omega_2$. More generally, one has
\begin{equation}\label{eq:permutation}
    B_{\ldots,n_i,\ldots,n_j,\ldots}=B_{\ldots,n_j,\ldots,n_i,\ldots}(g_i\rightleftharpoons g_j,\omega_i\rightleftharpoons\omega_j)
\end{equation}

Apart from the observation on the series, one can also look into the recurrence. Notice that $B_{\boldsymbol{n}}$ is given by Eq. (\ref{recurrence-1}), and exchanging $g$ and $\omega$ will affect the value of $B$. Taking the exchange of the first and the second modes as an example, Eq. (\ref{recurrence-1}) changes to 
\begin{equation}
\begin{split}
&g_2n_1B_{n_1,n_2-1,\ldots}+g_1n_2B_{n_1-1,n_2,\ldots}+\cdots\\
      =&f_{n_1-1,n_2-1,\ldots}B_{n_1-1,n_2-1,\ldots}-g_2B_{n_1-2,n_2-1,\ldots}-g_1B_{n_1-1,n_2-2,\ldots}-\cdots.
\end{split} 
\end{equation}
Also, by renaming the subscripts, one has
\begin{equation}\label{recurrence-2}
\begin{split}
&g_1n_1B_{n_2-1,n_1,\ldots}+g_2n_2B_{n_2,n_1-1,\ldots}+\cdots\\
      =&f_{n_2-1,n_1-1,\ldots}B_{n_2-1,n_1-1,\ldots}-g_1B_{n_2-1,n_1-2,\ldots}-g_2B_{n_2-2,n_1-1,\ldots}-\cdots.
\end{split} 
\end{equation}
From Eq. (\ref{eq:fun}), it can be noticed that $f_{n_1-1,n_2-1}=f_{n_2-1,n_1-1}(\omega_1\rightleftharpoons\omega_2)$. Then, compared with the original Eq. (\ref{recurrence-1}), the solutions to Eq. (\ref{recurrence-2}) have a correspondence relation with those of the Eq. (\ref{recurrence-1}),
\begin{equation}
\begin{split}
&B_{n_2-1,n_1,\ldots}\rightarrow B_{n_1,n_2-1,\ldots}\\
&B_{n_2,n_1-1,\ldots}\rightarrow B_{n_1-1,n_2,\ldots}\\
&B_{n_2,n_1,\ldots}\rightarrow B_{n_1,n_2,\ldots}\\
&\cdots.
\end{split} 
\end{equation}
Therefore, the solutions to Eq. (\ref{recurrence-1}) naturally satisfy the property of Eq. (\ref{eq:permutation}). However, as previously obtained, if one of the subscripts in $B_{\boldsymbol{n}}$ is zero, a boundary-type equation of it, i.e. $0\cdot B_{\boldsymbol{n}}=0$, can be found by properly setting the subscript of Eq. (\ref{recurrence-1}). This indicates that $B_{\boldsymbol{n}}$ with at least one zero subscript can be set arbitrarily. Combining the results, one can conclude that the solutions of $B_{\boldsymbol{n}}$ will satisfy the symmetry no matter what kind of boundary conditions is chosen, but the boundary conditions themselves do not follow the symmetry.

Now, we go back to equation $\sum_{j=1}^Ng_jB_{\boldsymbol{e}_j}=f_{\boldsymbol{0}}$. Notice that each $B_{\boldsymbol{e}_j}$ has a boundary-type equation. Therefore, $B_{\boldsymbol{e}_j}$ can be solved by the set of equations composed of the particular boundaries and  $\sum_{j=1}^Ng_jB_{\boldsymbol{e}_j}=f_{\boldsymbol{0}}$. For example, one can set $B_{1,0,\ldots}=B_{0,1,\ldots}=\ldots=B_{\ldots,1,0}=1$ and solve $B_{\ldots,0,1}$ by $\sum_{j=1}^Ng_jB_{\boldsymbol{e}_j}=f_{\boldsymbol{0}}$. However, remember that all $B$s have to satisfy the exchange rules due to the symmetry requirements. So, if one would like to obtain the $G$-function which happens to be the one under certain exchange of $g$ and $\omega$, the corresponding boundaries must also be exchanged. Otherwise, the symmetry will not remain in the spectrum. A convenient way to avoid the change of boundaries is to introduce tighter conditions for them. Back to the example of $B_{\boldsymbol{e}_j}$, one can set that all of them are equal, and then $B_{\boldsymbol{e}_j}=f_{\boldsymbol{0}}/\sum_{j=1}^Ng_j$. 

More generally, for each $p\geq N$, a set of linear equations can be obtained. As discussed previously, each set of equations is obtained by substituting $\boldsymbol{n}$ under a certain $p$. Therefore, as a recurrence, the variables are those with larger subscripts (left-hand side of Eq. (\ref{recurrence-1})) for a certain $p$. The total number of them equals the number of $N$-tuples $\boldsymbol{n}$ under that the total bosonic mode number is $p-(N-1)$. This is due to the form of the subscripts on the left-hand side of Eq. (\ref{recurrence-1}), $\boldsymbol{n-1}+\boldsymbol{e}_j$. In particular, the number of variables is
\begin{equation}
    \binom{p}{N-1}.
\end{equation}
The number of boundary-type equations ($0\cdot B=0$) equals the number of $N$-tuples with at least one zero. Therefore, the number of effective equations equals the number of $N$-tuples without zeros. In such a case, the $N$-tuples can be expressed by $\boldsymbol{n}'+\boldsymbol{1}$, where $\boldsymbol{1}$ is the all-one $N$-tuple and $\boldsymbol{n}'$ is the $N$-tuples given total bosonic numbers $p-N$. Therefore, the number of effective equations is given by 
\begin{equation}
   \binom{p-1}{N-1}.
\end{equation}
Using the property of binomial coefficients, it can be found that 
\begin{equation}
   \binom{p-1}{N-1}<\binom{p}{N-1}.
\end{equation}
This means that the number of equations obtained under a particular $p$ is less than the number of variables, making the equation set underdetermined. However, because of the boundaries, the number of variables can be largely reduced. In fact, some $N$-tuples under total boson number $p$ give the same boundary-type equations. Therefore, the number of variables that has at least one zero-subscript (the number of $N$-tuples $\boldsymbol{n}$ under $p-(N-1)$ with at least one zero component) will contribute to the true number of variables. Extending the symmetric boundary conditions of $B_{\boldsymbol{e}_j}$, for given $p$, one can set the following rules at most
\begin{equation}
\begin{split}
&B_{p-N+1,0,\ldots}=B_{0,p-N+1,\ldots}=\cdots=B_{\ldots,p-N+1},\\
&B_{p-N,1,0,\ldots}=B_{1,p-N,0\ldots}=\cdots=B_{\ldots,1,p-N},\\
&\cdots.
\end{split}
\end{equation}
Therefore, the total number of variables can be reduced to the number of $N$-tuples with no zeros plus the number of the above rules, under $p-N+1$. The number of the above rules equals the total number of ways to divide $p-N+1$ into the summation of less than $N-1$ positive integers. Hence, the number of variables after introducing the exchange rules is 
\begin{equation}
    \binom{p-N-2}{N-1}+\sum_{k=1}^{N-1}\binom{p-N}{k-1}
\end{equation}
Similarly, an inequality can be found by the binomial coefficients,
\begin{equation}
    \binom{p-1}{N-1}>\binom{p-N-2}{N-1}+\sum_{k=1}^{N-1}\binom{p-N}{k-1},
\end{equation}
for $p>N$ (remember that the case when $p<N$ has been discussed). The inequity shows that, by introducing the exchanging rules, the number of variables can be reduced to be lower than the number of equations. Therefore, for a given $p$, by introducing a proper number of rules based on permutation symmetry, a linear set of equations for $B_{\boldsymbol{n-1}+\boldsymbol{e}_j}$ is always solvable, which ultimately leads to the integrability of the whole model.

\section{IV. The example of the two-mode case}
We briefly illustrate the formulas for the two-mode case. In such a case, the Hamiltonian is given by 
\begin{equation}
\begin{split}
    H_2=\omega_1 a_1^{\dagger}a_1+g_1\sigma_x(a_1^{\dagger}+a_1)+\omega_2a_2^{\dagger}a_2+g_2\sigma_x(a_2^{\dagger}+a_2)+\Delta\sigma_z.
\end{split} 
\end{equation}
Using the method in the previous section, one obtains the $G$-function for the two-mode case,
\begin{equation}
\begin{split}
    &G_2^{\pm}(X)=\sum_{n_1,n_2=0}^\infty B_{n_1,n_2}\left[1\mp\frac{\Delta}{X-(n_1\omega_1+n_2\omega_2)}\right]\left(\frac{g_1}{\omega_1}\right)^{n_1}\left(\frac{g_2}{\omega_2}\right)^{n_2}, 
\end{split}
\end{equation}
where $X=E+g_1^2/\omega_1+g_2^2/\omega_2$. $B_{n_1,n_2}$ is given by 
\begin{equation}
\begin{split}
    g_1n_1B_{n_1,n_2-1}+g_2n_2B_{n_1-1,n_2}=f_{n_1-1,n_2-1}B_{n_1-1,n_2-1}-g_1B_{n_1-2,n_2-1}
    -g_2B_{n_1-1,n_2-2}.    
\end{split}
\end{equation}
As presented in the previous section, sets of equations for $B_{n_1,n_2}$ can be obtained by evaluating $n_1$ and $n_2$ and solved. The evaluation can be performed by setting the integer $p=n_1+n_2$ from 1 to $\infty$. In this case, it is worth noting that one can merely set the boundary $B_{n',0}=B_{0,n'}$ to obtain fully deterministic sets of equations.  


\begin{thebibliography}{99}
\bibitem{Leggett1987}
A. J. Leggett, S. Chakravarty, A. T. Dorsey, M. P. A. Fisher, A. Garg, and W. Zwerger, \rmp{59}, 1 (1987).

\bibitem{Rabi1936}
I. I. Rabi, \pr{49}, 324 (1936); {\bf 51}, 652 (1937).

\bibitem{Malekakhlagh2019}
M. Malekakhlagh and A, W. Rodriguez, \prl{122}, 043601 (2019)

\bibitem{Lu2023}
X. Lu, H. Li, J.-K. Shi, L.-B. Fan, V. Mangazeev, Z.-M. Li, and M. T. Batchelo, \pra{108}, 053712 (2023)

\bibitem{Brune1996}
M. Brune, F. Schmidt-Kaler, A. Maali, J. Dreyer, E. Hagley, J. M. Raimond, and S. Haroche, \prl{76}, 1800 (1996).

\bibitem{Raimond2001}
J. M. Raimond, M. Brune, and S. Haroche, \rmp{73}, 565 (2001).

\bibitem{Mabuchi2002}
H. Mabuchi and A. Doherty, \sci{298}, 1372 (2002).

\bibitem{Walther2006}
H. Walther, B. T. Varcoe, B.-G. Englert, and T. Becker, Rep. Prog. Phys. {\bf 69}, 1325 (2006).

\bibitem{Wallraff2004} 
A. Wallraff, D. I. Schuster, A. Blais, L. Frunzio, R.-S. Huang, J. Majer, S. Kumar, S. M. Girvin, and R. J. Schoelkopf, Nature (London) {\bf 431}, 162 (2004).

\bibitem{Blais2004}
A. Blais, R.-S. Huang, A. Wallraff, S. M. Girvin, and R. J. Schoelkopf, \pra{69}, 062320 (2004).

\bibitem{Chiorescu2004} 
I. Chiorescu, P. Bertet, K. Semba, Y. Nakamura, C. Harmans, and J. Mooij, Nature (London) {\bf 431}, 159 (2004).

\bibitem{Schuster2007}
D. I. Schuster, A. A. Houck, J. A. Schreier, A. Wallraff, J. M. Gambetta, A. Blais, L. Frunzio, J. Majer, B. Johnson, M. H. Devoret, S. M. Girvin, and R. J. Schoelkopf, Nature (London) {\bf 445}, 515 (2007).

\bibitem{Clarke2008}
J. Clarke and F. K. Wilhelm, Nature (London) 453, 1031 (2008).

\bibitem{Hofheinz2009} M. Hofheinz, H. Wang, M. Ansmann, R. C. Bialczak, E. Lucero, M. Neeley, A. O’connell, D. Sank, J. Wenner, J.M. Martinis et al., Nature (London) {\bf 459}, 546 (2009).

\bibitem{Leibfried2003} D. Leibfried, R. Blatt, C. Monroe, and D. Wineland, \rmp{75}, 281 (2003).

\bibitem{Pedernales2015} J. Pedernales, I. Lizuain, S. Felicetti, G. Romero, L. Lamata,
and E. Solano, Sci. Rep. {\bf 5}, 15472 (2015).

\bibitem{Hennessy2007}
K. Hennessy, A. Badolato, M. Winger, D. Gerace, M.
Atatüre, S. Gulde, S. F{\"a}lt, E. L. Hu, and A. Imamo{\v g}lu, Nature (London) {\bf 445}, 896 (2007).

\bibitem{Irish2003}
E. K. Irish and K. Schwab, \prb{68}, 155311 (2003).

\bibitem{Cleland2004} 
A. N. Cleland and M. R. Geller, \prl{93}, 070501 (2004).

\bibitem{LaHaye2009} 
M. LaHaye, J. Suh, P. Echternach, K. C. Schwab, and M. L. Roukes, Nature (London) {\bf 459}, 960 (2009).

\bibitem{Albert2012}
V. V. Albert, \prl{108}, 180401 (2012).

\bibitem{Irish2007}
E. K. Irish, \prl{99}, 173601 (2007).

\bibitem{Wu2002}
L.-A. Wu, M. S. Byrd, and D. A. Lidar, \prl{89}, 057904 (2002).

\bibitem{Crespi2012}
A. Crespi,1,2 S. Longhi,1,2 and R. Osellame, \prl{108}, 163601 (2012).

\bibitem{Wu2024}
Z. Wu, C. Hu, T. Wang, Y. Chen, Y. Li, L. Zhao, X.-Y. L{\"u}, and X. Peng, \prl{133}, 173602 (2024).

\bibitem{Jing2018}
J. Jing and L.-A. Wu, Sci. Rep. {\bf 8}, 1471 (2018).

\bibitem{Wu2002-1}
L.-A. Wu and D. A. Lidar, J. Math. Phys. {\bf 43}, 4506–4525 (2002)

\bibitem{Jaynes1963}
E. T. Jaynes and F.W. Cummings, Proc. IEEE {\bf 51}, 89 (1963).

\bibitem{Niemczyk2010}
T. Niemczyk, F. Deppe, H. Huebl, E. Menzel, F. Hocke, M. Schwarz, J. Garcia-Ripoll, D. Zueco, T. Hümmer, E. Solano, et al., Nat. Phys. {\bf 6}, 772 (2010).

\bibitem{FornDiaz2010} 
P. Forn-D{\'i}az, J. Lisenfeld, D. Marcos, J. J. García-Ripoll, E. Solano, C. J. P. M. Harmans, and J. E. Mooij, \prl{105}, 237001 (2010).

\bibitem{Yoshihara2017}
F. Yoshihara, T. Fuse, S. Ashhab, K. Kakuyanagi, S. Saito, and K. Semba, Nat. Phys. {\bf 13}, 44 (2017).

\bibitem{Albert2011}
V. V. Albert, G. D. Scholes, and P. Brumer, \pra{84}, 042110 (2011).

\bibitem{Braak2011}
D. Braak, \prl{107}, 100401 (2011).

\bibitem{Schweber1967} 
S. Schweber, Ann. Phys. (N.Y.) {\bf 41}, 205 (1967).

\bibitem{Swain1973} 
S. Swain, J. Phys. A {\bf 6}, 192, 1919 (1973).

\bibitem{Durst1986} 
C. Durst, E. Sigmund, P. Reineker, A. Scheuing, J. Phys. C: Solid State Phys. {\bf 19}, 2701 (1986)

\bibitem{Chen2012} Q.-H. Chen, C. Wang, S. He, T. Liu, and K.-L. Wang, \pra{86}, 023822 (2012).

\bibitem{Braak2013}
D. Braak, J. Phys. B: At. Mol. Opt. Phys. {\bf 46}, 224007 (2013).

\bibitem{Peng2013}
J. Peng, Z. Ren, G. Guo, G. Ju, and X. Guo, Eur. Phys. J. D {\bf 67}, 162 (2013).

\bibitem{Zhang2014}
Y.-Z. Zhang, Ann. Phys. {\bf 347}, 122-129 (2014).

\bibitem{Duan2015}
L. Duan, S. He, D. Braak, and Q.-H. Chen, \epl{112}, 34003 (2015).

\bibitem{He2015}
S. He, L. Duan and Q.-H. Chen, New J. Phys. {\bf 17}, 043033 (2015).

\bibitem{Wu2017}
L.-A. Wu, J. Phys. A: Math. Theor. {\bf 50}, 255204 (2017).

\bibitem{Xie2017}
Q. Xie, H. Zhong, M. T. Batchelor, and C. Lee, J. Phys. A: Math. Theor. {\bf 50}, 113001 (2017).

\bibitem{Cui2017}
S. Cui, J.-P. Cao, H. Fan, and L. Amico, J. Phys. A: Math. Theor. {\bf 50}, 204001 (2017).

\bibitem{Peng2017}
J. Peng, C. zheng, G. Guo, X. Guo, X. Zhang, C. Deng, G. Ju, Z. Ren, L. Lamata, and E. Solano, J. Phys. A: Math. Theor. {\bf 50}, 174003 (2017).

\bibitem{Wu2003}
L.-A. Wu, M. Guidry, Y. Sun, and C.-L. Wu, \prb{67}, 014515 (2003).

\bibitem{Wu2017-2}
L.-A. Wu, M. Murphy, and M. Guidry, \prb{95}, 115117 (2017).

\bibitem{Wu2021}
L.-A. Wu, Introduction to Exactly Solvable Models (2021), doi: 10.13140/RG.2.2.27415.29600.

\bibitem{Eberly1980}
J. H. Eberly, N. B. Narozhny, J. J. Sanchez-Mondragon, \prl{44}, p.1323 (1980). 

\bibitem{Vaaranta2025}
A. Vaaranta, M. Cattaneo, and P. Muratore-Ginanneschi, \pra{111}, 053717 (2025).

\bibitem{Chen2022}
L. Chen, X.-W. An, T.-H. Deng, and Z.-R. Zhong, Quant. Inf. Process {\bf 21}, 232 (2022).

\bibitem{Gerry1988}
C. C. Gerry, \pra{37}, 2683 (1988).

\bibitem{Valle2010}
E. del Valle, S. Zippilli, F. P. Laussy, A. Gonzalez-Tudela, G. Morigi, and C. Tejedor, \prb{81}, 035302 (2010).

\bibitem{Felicetti2015}
S. Felicetti, J. S. Pedernales, I. L. Egusquiza, G. Romero, L. Lamata, D. Braak, and E. Solano, \pra{92}, 033817 (2015).

\bibitem{Travenec2012}
I. Tra{\v v}enec, \pra{85}, 043805 (2012). 

\bibitem{Maciejewski}
A. J. Maciejewski, M. Przybylska, and T. Stachowiak, \pra{91}, 037801 (2015).

\bibitem{Zhang2015}
Y.-Z. Zhang, arXiv:1507.03863.

\bibitem{Bargmann1961}
V. Bargmann, Comm. Pure Appl. Math. {\bf 14}, 197 (1961).

\bibitem{Bargmann1962}
V. Bargmann, Proc. Natl. Acad. Sci. {\bf 48} (2), 199-204 (1962).

\bibitem{Vukics2018}
A. Vukics and P. Domokos, J. Russ. Laser Res. {\bf 39}, 353-359 (2018).

\bibitem{Chabaud2022}
U. Chabaud and S. Mehraban, Quantum {\bf 6}, 831 (2022).

\bibitem{Styan1973}
G. P. H. Styan, Linear Algebra Appl. {\bf 6}, 217 (1973).

\end{thebibliography}
\end{document}